\documentclass[twocolumn,aps,pra,showpacs,floatfix,superscriptaddress]{revtex4-1}
\usepackage{graphicx} 
\usepackage{amsmath}
\usepackage{bm}
\usepackage{color}
\usepackage{dcolumn}

\begin{document}
\newcommand{\NZIAS}{
Centre for Theoretical Chemistry and Physics,
The New Zealand Institute for Advanced Study,
Massey University Auckland, Private Bag 102904, 0745 Auckland, New Zealand
}

\title{Enhanced Sensitivity to the Time Variation of the Fine-Structure Constant and $m_p/m_e$ in Diatomic Molecules: A Closer Examination of Silicon Monobromide}

\author{K. Beloy}
\affiliation{\NZIAS}

\author{A. Borschevsky}
\affiliation{\NZIAS}

\author{P. Schwerdtfeger}
\affiliation{\NZIAS}

\author{V. V. Flambaum}
\affiliation{School of Physics, University of New South Wales, Sydney 2052, Australia}
\affiliation{\NZIAS}

\date{\today}

\begin{abstract}
Recently it was pointed out that transition frequencies in certain diatomic molecules have an enhanced sensitivity to variations in the fine-structure constant $\alpha$ and the proton-to-electron mass ratio $m_p/m_e$ due to a near cancellation between the fine-structure and vibrational interval in a ground electronic multiplet [V.~V.~Flambaum and M.~G.~Kozlov, Phys. Rev. Lett.~{\bf 99}, 150801 (2007)]. One such molecule possessing this favorable quality is silicon monobromide. Here we take a closer examination of SiBr as a candidate for detecting variations in $\alpha$ and $m_p/m_e$. We analyze the rovibronic spectrum by employing the most accurate experimental data available in the literature and perform \emph{ab initio} calculations to determine the precise dependence of the spectrum on variations in $\alpha$. Furthermore, we calculate the natural linewidths of the rovibronic levels, which place a fundamental limit on the accuracy to which variations may be determined.
\end{abstract}

\maketitle

\newcommand{\cmt}[1]{[\![#1]\!]}

\section{Introduction}

Theories unifying gravity with other interactions suggest the possibility of
spatial and temporal variation of fundamental physical constants (VFC), such as
the fine structure constant, $\alpha =e^{2}/\hbar c$, and the proton-to-electron
mass ratio, $\mu =m_{p}/m_{e}$ \cite{Uza03}. Search for such
variation has received considerable interest in recent years, and is being
conducted using a wide variety of methods \cite{Fla07,Fla08}.
Some major directions of this research include analysis of high resolution
spectroscopy of quasar absorption systems \cite{WebFlaChu99,WebMurFla01,MurWebFla07}%
, frequency comparison of atomic clocks over extended periods of time \cite%
{PreTjoMal95,ForAshBer07,SorBizAbg01}, and nuclear methods, including study of
nucleosynthesis, alpha and beta decay, and Oklo natural reactor \cite%
{OliPos02,OliPosQia02,LamTor04,FujIwaFuk00,DamDys96,DmiFlaWeb04}.

Precision molecular spectroscopy is a new and promising direction of search
for variation of fundamental constants. Molecular spectra are sensitive to
both $\mu$ and $\alpha$, and by measuring close lying levels great
enhancement of relative variation may be observed \cite%
{Fla08,ChiFlaKoz09,MurFlaMul08}. In particular, diatomic molecules that have
a near cancellation between hyperfine structure and rotational intervals or
between fine structure and vibrational intervals are of interest in the
context of such an enhancement. A number of such molecules have been proposed,
e.g. Cs$_{2}$ \cite{DemSaiSag08},
CaH, MgH, CaH$^{+}$ \cite{Kaj08,KajMor09}, Cl$_2^{+}$, IrC, HfF$^{+}$,
SiBr, LaS, LuO, and others \cite{FlaKoz07}.


In this paper, we conduct a detailed study of one of the molecular candidates suggested by Flambaum and Kozlov \cite{FlaKoz07}, namely silicon monobromide. To this end, it is useful to start by briefly recapitulating some of the main concepts put forth by these authors. We consider a diatomic molecule with an electronic ground state composed of a fine structure multiplet. Taking the vibrational energy spacing of the multiplet as $\omega_e$ in the harmonic approximation, and the fine structure (spin-orbit) energy spacing between two multiplet states as $\omega_f$, the energy associated with a transition between the multiplet states reads
\begin{equation*}
\omega=\omega_f-v\omega_e,
\end{equation*}
where $v$ represents the change in the number of vibrational quanta for the transition.

The fine structure and vibrational energies have different sensitivities to variations in $\alpha$ and $\mu$. In particular, $\omega_f$ is sensitive to variations in the fine structure constant, scaling as $\omega_f\sim\alpha^2$, while being almost insensitive to variations in $\mu$. On the other hand, $\omega_e$ is sensitive to variations in the proton-to-electron mass ratio, scaling as $\omega_e\sim\mu^{-1/2}$, while being insensitive to variations in $\alpha$. It follows that $\omega$ is sensitive to variations in both $\alpha$ and $\mu$, with a corresponding variation for fractional variations in $\alpha$ and $\mu$ given by
\begin{equation*}
\delta\omega=2\omega_f\frac{\delta\alpha}{\alpha}+\frac{v}{2}\omega_e\frac{\delta\mu}{\mu}.
\end{equation*}

For a number of molecules there exist transitions having a near cancellation between fine structure and vibrational energies, i.e., $\omega_f\approx v\omega_e$. In such cases, the fractional variation of $\omega$ may then be written
\begin{equation*}
\frac{\delta\omega}{\omega}\approx K\left(2\frac{\delta\alpha}{\alpha}+\frac{1}{2}\frac{\delta\mu}{\mu}\right),
\end{equation*}
where $K=\omega_f/\omega$ is an enhancement factor. Large values of $K$ are suggestive of favorable cases for experimentally detecting a signal from variations in $\alpha$ or $\mu$. As discussed in Ref.~\cite{FlaKoz07}, however, it is also necessary to consider the size of the absolute shift $\delta\omega$ and compare this to experimental limitations on measuring $\omega$ itself; one such notable limitation is the natural linewidth and intensity of the transition.

The diatomic molecule SiBr has a $^2\Pi_r$ electronic ground state with fine structure and vibrational spacing similar to about 1 cm$^{-1}$ ($\omega_f\approx \omega_e\approx420~\mathrm{cm}^{-1}$). This is comparable to the rotational constant $B_e$, and thus $\omega$ may be reduced further by a suitable choice of rotational levels. In this paper 
we examine the rovibronic spectrum of SiBr by employing the most accurate experimental spectroscopic data for SiBr available in the literature, namely that of Bosser~\emph{et al.}~\cite{BosLebRos81}. Furthermore we perform \emph{ab initio} molecular calculations with the purpose of i) determining the precise dependence of the spectrum on $\alpha$, and ii) obtaining values for the natural linewidths of the pertinent levels.
As in Ref.~\cite{FlaKoz07}, we still conclude that dedicated measurements are required to determine precise values of transition frequencies and find the best transitions for the search of VFC; the aim of this work to entice experimental progress in this direction.

At the risk of being overly prudent, we discuss our convention used throughout concerning units, applicable to the above expressions as well. We choose to work with atomic units ($\hbar=e=m_e=1$), and thus an expression such as $\delta X$ indicates a variation in $X$ when expressed in atomic units (this is not a trivial remark: for instance, when expressed in atomic units the speed of light $c=1/\alpha$ certainly varies with a variation of $\alpha$; however \emph{by definition} the speed of light does not change if expressed in SI units). Throughout this paper we will find it useful to express energy values in the spectroscopically familiar units of cm$^{-1}$; one should interpret this merely as a conversion from the atomic unit of energy, $1~\mathrm{a.u.}=2.19474625\times10^5~\mathrm{cm}^{-1}$. In the end we will only be concerned with variations of dimensionless quantities \cite{Uza03}, such as the ratio of two frequencies, and for these expressions ambiguity surrounding units is non-existent.

\section{Rovibronic energy levels in Hund's case $a$ diatomics}
\label{Sec:rovibronicspectra}
We consider an electronic multiplet of a diatomic molecule which is categorically described by Hund's case $a$ \cite{Kro75}. In Hund's case $a$, the electronic orbital angular momentum $\mathbf{L}$ is strongly coupled to the internuclear axis (chosen to be the $z$-axis in a molecule-fixed frame), which is to say that $\Lambda$, the eigenvalue of $L_z$, remains a good quantum number. Furthermore, the spin angular momentum $\mathbf{S}$ is strongly coupled to the internuclear axis by way of the spin-orbit interaction, and thus $\Sigma$, the eigenvalue of $S_z$, also remains a good quantum number.

Initially we neglect the spin-orbit interaction, in which case we may write the vibronic energies $T_{ev}$ of a given electronic multiplet in terms of conventional spectroscopic constants,
\begin{equation}
T_{ev}=T_e+\omega_e\left(v+\frac{1}{2}\right)-\omega_ex_e\left(v+\frac{1}{2}\right)^2,
\label{Eq:Tev}
\end{equation}
where $v$ is the vibrational quantum number and terms beyond second order in $(v+\frac{1}{2})$ are omitted. The constant $T_e$ is the energy relative to the ground state multiplet; as we will only be concerned with the ground state multiplet, we may set $T_e=0$. Constants $\omega_e$ and $\omega_ex_e$ represent the harmonic vibrational energy and the first correction due to anharmocity, respectively.

Next we consider the effective spin-orbit interaction. As $\mathbf{L}$ is strongly coupled to the internuclear ($z$) axis, the spin-orbit interaction takes the simple form \cite{BroWat77}
\begin{equation}
H_\mathrm{so}=A_vL_zS_z.
\label{Eq:Hso}
\end{equation}
The spin-orbit factor $A_v$ here depends on the vibrational state and to the first order in $(v+\frac{1}{2})$ may be written as
\begin{equation}
A_v=A_e-\alpha_{Ae}\left(v+\frac{1}{2}\right).
\label{Eq:Av}
\end{equation}

Finally, we consider the energy associated with rotation, taking the effective rotational Hamiltonian for the Hund's intermediate case $a-b$ as in Ref.~\cite{BroWat77},
\begin{equation}
H_\mathrm{rot}=B_vN^2,
\label{Eq:Hrot}
\end{equation}
where $\mathbf{N}=\mathbf{J}-\mathbf{S}$, and $\mathbf{J}$ being the total angular momentum excluding nuclear spin. We now introduce the operators $J^\pm=J_x\pm iJ_y$, and similar for $S^\pm$, where $x$ and $y$ correspond to the molecule-fixed axes perpendicular to the internuclear axis. With these operators we expand $N^2$ as
\begin{eqnarray}
N^2&=&J^2+S^2-2\mathbf{J}\cdot\mathbf{S}\nonumber\\
&=&J^2+S^2-2S_z^2-2L_zS_z-\left(J^+S^-+J^-S^+\right),~~~
\label{Eq:N2}
\end{eqnarray}
where we have used $J_z-L_z-S_z=0$ with the physical reasoning that the molecule rotates about an axis perpendicular to the internuclear axis. In the expressions to follow, we neglect the small $v$-dependence of $B_v$ and use $B_v=B_e$. For the Hund's case $a$ limit, where $\Sigma$ is assumed to be a good quantum number, the term in parenthesis in Eq.~(\ref{Eq:N2}) may be dropped as it involves the raising and lowering operators $S^\pm$.

We now consider the energy levels specific to a $\Pi$-doublet, such as the ground electronic state of SiBr. The appropriate basis for Hund's case $a$ is $|JM,\Lambda\Sigma\rangle$, where $M$ is the eigenvalue of $J_Z$, $Z$ being a space-fixed axis. In this basis, the doubly-degenerate $^2\Pi_{1/2}$ state is represented by $|JM,\pm1,\mp\frac{1}{2}\rangle$ and the doubly-degenerate $^2\Pi_{3/2}$ state represented by $|JM,\pm1,\pm\frac{1}{2}\rangle$. The angular momentum quantum number $J$ is necessarily a half-integer, with limitations $J\ge1/2$ and $J\ge3/2$ for the $^2\Pi_{1/2}$ and $^2\Pi_{3/2}$ states, respectively.
In terms of the spectroscopic constants introduced above, the energy levels are then given by
\begin{eqnarray}
E_{vJ}&=&\pm \frac{1}{2}(A_e-2B_e)
+\left(\omega_e\mp\frac{1}{2}\alpha_{Ae}\right)\left(v+\frac{1}{2}\right)\nonumber\\
&&-\omega_ex_e\left(v+\frac{1}{2}\right)^2
+B_e\left(J+\frac{1}{2}\right)^2,
\label{Eq:EvJ}
\end{eqnarray}
where the top (bottom) sign corresponds to the $^2\Pi_{3/2}$ ($^2\Pi_{1/2}$) levels. As discussed in Ref.~\cite{BroWat77}, $\alpha_{Ae}$ may be regarded as the difference in the harmonic vibrational energies of the doublet levels when considered independently, i.e., $\alpha_{Ae}=\omega_e^{(1/2)}-\omega_e^{(3/2)}$; this interpretation is consistent with Eq.~(\ref{Eq:EvJ}).

It will be useful to separate the energy given by Eq.~(\ref{Eq:EvJ}) into $J$-independent and $J$-dependent parts,
\begin{equation*}
E_{vJ}=G_v+F_J,
\end{equation*}
where $F_J=B_e\left(J+\frac{1}{2}\right)^2$. We will refer to these as the vibronic and rotational energies, respectively. Note that with this nomenclature the energy contribution $\mp B_e$ of Eq.~(\ref{Eq:EvJ}) is associated with the vibronic energy, despite arising from the rotational Hamiltonian, Eq.~(\ref{Eq:Hrot}). We further note that with this choice of separation, only the expression for the vibronic part depends on the particular doublet state ($^2\Pi_{3/2}$ or $^2\Pi_{1/2}$). The total energies, $E_{vJ}$ will be referred to as the rovibronic energies.

It is briefly noted here that certain additional terms, such as those associated with the spin-rotation interaction,
lambda-doubling, and the hyperfine interaction, have intentionally been neglected in this section. These contributions are generally small, though for large $J$ some of these terms may have a sizable effect.

\section{Rovibronic energy levels of silicon monobromide}
\begin{figure}[ht]
\begin{center}
\includegraphics*[scale=0.51]{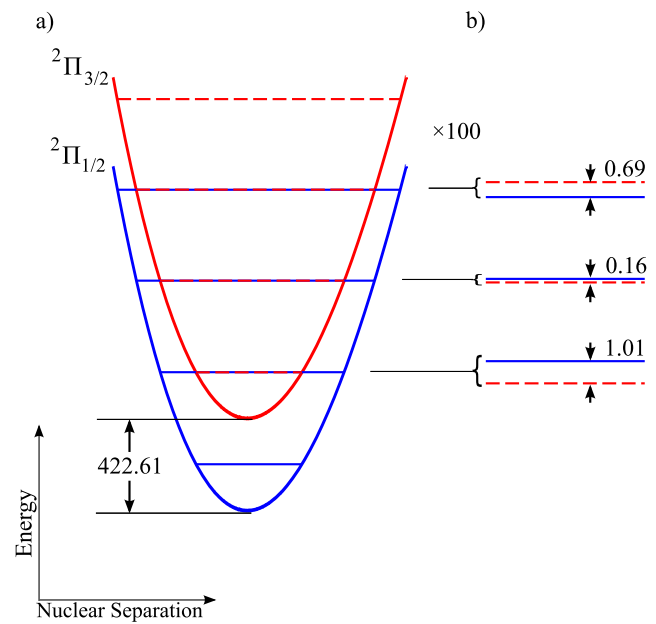}
\end{center}
\caption{(Color online) Potential energy curves and the lowest few vibronic energy levels for the $X~^2\Pi_r$ ground doublet of $^{28}$Si$^{79}$Br. (a) The lower blue and upper red curves represent the potential energy curves for the $^2\Pi_{1/2}$ and $^2\Pi_{3/2}$ states, respectively, with the  solid blue and dashed red horizontal lines illustrating the corresponding vibronic energy levels. (b) The energy separation between the quasi-degenerate vibronic levels is magnified by a factor of 100. All energy differences are in cm$^{-1}$.}
\label{Fig:potcurve}
\end{figure}

The ground electronic multiplet, $X~^2\Pi_r$, of SiBr falls into the category of Hund's case $a$. Accurate spectroscopic constants for this doublet have been experimentally determined by Bosser \emph{et al.}~\cite{BosLebRos81} and are presented in Table \ref{Tab:specconsts}. Figure~\ref{Fig:potcurve}(a) illustrates the potential energy curves for the $^2\Pi_{1/2}$ and $^2\Pi_{3/2}$ states near minima, based on the data for isotopic species $^{28}$Si$^{79}$Br; also displayed are the lowest few vibronic energy levels. Due to similarity in the magnitude of the $A_e$ and $\omega_e$ constants, the $G_{v}^{(1/2)}$ level is quasi-degenerate with the $G_{v-1}^{(3/2)}$ level for $v=1,2,\dots$. Figure~\ref{Fig:potcurve}(b) provides a magnification of the energy separation between these quasi-degenerate levels.

\begin{table}[ht]
\begin{center}
\caption{Spectroscopic constants for the $X~^2\Pi_r$ ground doublet of SiBr. Theoretical values for $^{28}$Si$^{79}$Br are calculated using the relativistic Fock space coupled cluster approach, described in Section V. Experimental data for isotope $^{28}$Si$^{79}$Br are from Ref.~\cite{BosLebRos81}, whereas data for isotope $^{28}$Si$^{81}$Br are inferred using the appropriate dependence on reduced mass per spectroscopic constant along with the ratio of reduced masses of the two isotopic species, 1.0065044. All values in the table are in cm$^{-1}$.
\label{Tab:specconsts}}
\begin{tabular}[c]{lddd}%
\hline\hline
\multicolumn{1}{l}{} &
\multicolumn{1}{c}{\quad$^{28}$Si$^{79}$Br\quad} &
\multicolumn{1}{c}{\quad$^{28}$Si$^{79}$Br\quad} &
\multicolumn{1}{c}{\quad$^{28}$Si$^{81}$Br\quad} \\
\multicolumn{1}{l}{Constant} &
\multicolumn{1}{c}{theor.} &
\multicolumn{1}{c}{expt.} &
\multicolumn{1}{c}{expt.} \\
\hline
$A_e$        	& 419.54	& 422.61    & 422.61  \\
$\omega_e$   	& 424.35	& 424.14    & 422.77  \\
$\omega_ex_e$	&   1.32	& 1.41      & 1.40    \\
$\alpha_{Ae}$	&   2.26	& 1.97      & 1.96    \\
$B_e$        	&   0.1634	& 0.1671    & 0.1660  \\
\hline\hline
\end{tabular}
\end{center}
\end{table}

Before proceeding we briefly discuss the accuracy of the experimental data in Table~\ref{Tab:specconsts}. Explicit uncertainties are not provided for the constants, but indications from Ref.~\cite{BosLebRos81} are that the data are likely to be accurate to $\sim\!0.1~\mathrm{cm}^{-1}$. For the illustrative purposes of this section we will treat the data as exact; for deviations on the order of $\sim\!0.1~\mathrm{cm}^{-1}$ the important qualitative features of the spectrum remain, and only minor modifications would be necessary.

As mentioned in the Introduction, we are interested in transitions having large enhancement factors, namely the transitions between the quasi-degenerate vibronic levels.  We define the small energy difference between quasi-degenerate vibronic levels as
\begin{eqnarray}
\Delta G_{v}&\equiv&
G_{v-1}^{(3/2)}-G_{v}^{(1/2)}\nonumber\\
&=&A_e-\omega_e-2B_e+v\left(2\omega_ex_e-\alpha_{Ae}\right).
\label{Eq:DGv}
\end{eqnarray}
(It is noted that $\Delta G_{v}$ as defined here is not related to the usual spectroscopic $\Delta G_{1/2}$, $\Delta G_{3/2}$, $\dots$) For the isotope $^{28}$Si$^{79}$Br this reduces to
\begin{equation*}
\Delta G_{v}=\left(-1.86+0.85v\right)~\mathrm{cm}^{-1}.
\end{equation*}
An interesting property of the $^{28}$Si$^{79}$Br vibronic spectrum is that $\Delta G_{v}$ is negative for $v=1,2$ and positive for $v\ge3$. This is plainly seen in Figure~\ref{Fig:potcurve}(b), where for the quasi-degenerate levels described by $v=1,2$ the $G_{v-1}^{(3/2)}$ energy (dashed red line) is below the $G_{v}^{(1/2)}$ energy (solid blue line), whereas the order is inverted for $v\ge3$. This inversion arises due to the anharmocity of the potentials, $\omega_ex_e$.

We now turn our attention to rotational energies. With our choice of separation for ``vibronic'' and ``rotational'' contributions to the total energy, the rotational energies are given by the same expression for both doublet states, i.e., $F_J^{(1/2)}=F_J^{(3/2)}=B_e(J+\frac{1}{2})^2$.  We will concern ourselves only with single-photon transitions, from which the angular momentum restriction $\Delta J=0,\pm1$ follows. For $\Delta J=0$ transitions, there is no change in rotational energy, and the corresponding measured transition lines for all $J$ may be blended, limiting the accuracy. For this reason, we focus on transitions with $\Delta J=\pm1$. We define the difference in rotational energy encompassing both of these cases as
\begin{equation}
\Delta F_{J}^\pm\equiv
F_{J\pm1}^{(3/2)}-F_{J}^{(1/2)}=
\left\{\begin{array}{l}
2B_e\left(J+1\right) \\
-2B_eJ
\end{array}\right.,
\label{Eq:DFJ}
\end{equation}
where $\Delta F_{J}^+$ and $\Delta F_{J}^-$ are restricted by $J\ge1/2$ and $J\ge5/2$, respectively. We note that $\Delta F_{J}^+$ is necessarily positive, whereas $\Delta F_{J}^-$ is necessarily negative.

The experimentally observable quantity is the energy difference between two rovibronic levels; we define the energy difference between pertinent rovibronic levels as
\begin{equation}
\Delta E_{vJ}^\pm\equiv
E_{v-1,J\pm1}^{(3/2)}-E_{vJ}^{(1/2)}
=\Delta G_{v}+\Delta F_{J}^\pm.
\label{Eq:DEvJ}
\end{equation}
To continue with our strategy of finding the largest enhancement factors, we look for specific transitions in which
\begin{equation}
\Delta E_{vJ}^\pm=\Delta G_{v}+\Delta F_{J}^\pm\approx0.
\label{Eq:0tran}
\end{equation}
As an example, we take the $v=1$ vibronic energy difference of $^{28}$Si$^{79}$Br, $\Delta G_{1}=-1.01~\mathrm{cm}^{-1}$. As $\Delta G_{1}$ is negative, we require $\Delta F_{J}^+$ in Eq.~(\ref{Eq:0tran}) and may subsequently solve for $J$,
\begin{equation*}
J\approx-\frac{\Delta G_1}{2B_e}-1=2.02,
\end{equation*}
which indicates that two appropriate choices for $J$ are $J=3/2$ and $J=5/2$, with corresponding values of $\Delta E_{vJ}^+$
\begin{eqnarray}
\Delta E_{1,3/2}^+&=&-0.18~\mathrm{cm}^{-1},\nonumber\\
\Delta E_{1,5/2}^+&=&+0.16~\mathrm{cm}^{-1}.
\label{Eq:DEchoice}
\end{eqnarray}

\begin{figure}[ht]
\begin{center}
\includegraphics*[scale=0.52]{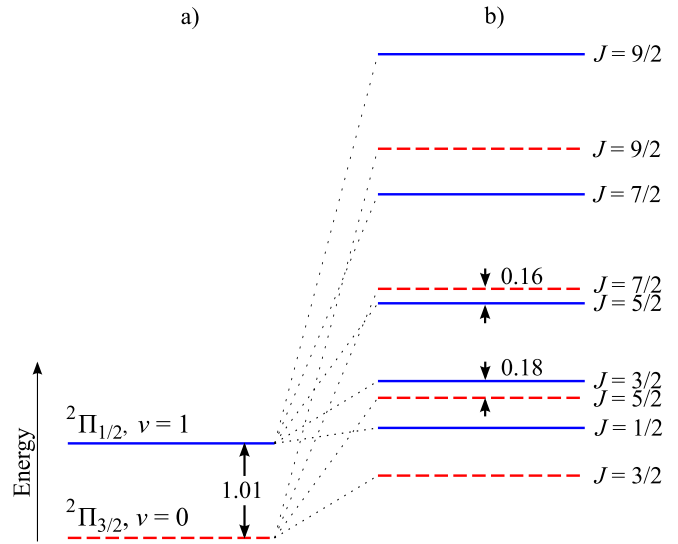}
\end{center}
\caption{(Color online) Rovibronic levels of $^{28}$Si$^{79}$Br associated with the $^2\Pi_{1/2},~v=1$ (solid blue lines) and $^2\Pi_{3/2},~v=0$ (dashed red lines) quasi-degenerate vibronic levels. Displayed are (a) the two vibronic levels and (b) all rovibronic levels for up to $J=9/2$. All energy differences are in cm$^{-1}$.}
\label{Fig:rovibronic}
\end{figure}

Figure~\ref{Fig:rovibronic} displays the rovibronic spectrum arising from the $G_{1}^{(1/2)}$ ($^2\Pi_{1/2},~v=1$) and $G_{0}^{(3/2)}$ ($^2\Pi_{3/2},~v=0$) quasi-degenerate vibronic levels for $^{28}$Si$^{79}$Br. The energy differences corresponding to $\Delta E_{1,3/2}^+$ and $\Delta E_{1,5/2}^+$ are also displayed. One can resolve the absolute values appearing in Figure~\ref{Fig:rovibronic} with the signed values appearing in Eq.~(\ref{Eq:DEchoice}) by noting that if the $E_{v-1,J\pm1}^{(3/2)}$ (red dashed line) is above the $E_{vJ}^{(1/2)}$ level (solid blue line), then the sign of $\Delta E_{vJ}^\pm$ is positive; if the order of the levels is opposite, the sign of $\Delta E_{vJ}^\pm$ is negative.
The particular sign of $\Delta E_{vJ}^\pm$ is important from the viewpoint of variations of $\Delta E_{vJ}^\pm$ with respect to variations of $\alpha$ and $\mu$, as discussed in the following section.

\section{Variations of the rovibronic transition frequencies with respect to variations of $\alpha$ and $\mu$}
In this section we consider variation of the energy difference $\Delta E_{vJ}^\pm$ with respect to variations of $\alpha$ and $\mu$. The constants $A_e$ and $\omega_e$ are orders of magnitude larger than the other spectroscopic constants used to describe $\Delta E_{vJ}^\pm$; furthermore $A_e$ is only sensitive to variations in $\alpha$ while $\omega_e$ is only sensitive to variations in $\mu$. Consequently, to a first approximation we can estimate the variation $\delta\left(\Delta E_{vJ}^\pm\right)$ by variations of $A_e$ and $\omega_e$,
\begin{equation*}
\delta\left(\Delta E_{vJ}^\pm\right)\cong\delta A_e-\delta\omega_e.
\end{equation*}

The spin-orbit constant $A_e$ embodies the major relativistic correction to the energy spectrum of the doublet and to the lowest order scales as $\alpha^2$. Thus, if we assume higher-order relativistic corrections to be negligible, we may write
\begin{equation*}
\delta A_e=2A_e\frac{\delta\alpha}{\alpha}.
\end{equation*}

The harmonic vibrational energy $\omega_e$ is insensitive to relativistic corrections, though it is proportional to $M_\mathrm{red}^{-1/2}$, where $M_\mathrm{red}$ is the reduced nuclear mass, and as such is sensitive to $\mu$. The proton and neutron masses, as well as nuclear binding energies, are all proportional to the quantum chromodynamics scale $\Lambda_\mathrm{QCD}$ (see, e.g., Refs.~\cite{FlaShu02,FlaShu03}). It follows that the nuclear masses and, further, the reduced nuclear mass are also proportional to $\Lambda_\mathrm{QCD}$. We conclude that $\delta M_\mathrm{red}/M_\mathrm{red}=\delta m_p/m_p=\delta \mu/\mu$, where the last equality holds for atomic units. The constant $\omega_e$ then varies with $\mu$ as
\begin{equation*}
\delta \omega_e=-\frac{1}{2}\omega_e\frac{\delta\mu}{\mu}.
\end{equation*}

Combining the above equations yields
\begin{eqnarray}
\delta\left(\Delta E_{vJ}^\pm\right)&\cong&2A_e\frac{\delta\alpha}{\alpha}+\frac{1}{2}\omega_e\frac{\delta\mu}{\mu}\nonumber\\
&\cong&2A_e\left(\frac{\delta\alpha}{\alpha}+\frac{1}{4}\frac{\delta\mu}{\mu}\right),
\label{Eq:dwest}
\end{eqnarray}
where in the last expression we used the fact that $A_e\cong\omega_e$. Evidently the transitions $\Delta E_{vJ}^\pm$ are sensitive to variations in the combined constant $\alpha\mu^{1/4}$ (the term in parentheses is equivalent to the fractional variation $\delta\rho/\rho$, where $\rho=\alpha\mu^{1/4}$).

As discussed in the Introduction, the variation $\delta\left(\Delta E_{vJ}^\pm\right)$ is dependent on our choice of unit system, namely atomic units. To remove dependence on the unit system we consider variations of dimensionless quantities, such as the ratio of two transition energies.
In the present set-up, we consider two transition energies within the same doublet rovibronic spectrum. Equating $\omega_1$ and $\omega_2$ to two separate transition energies $\Delta E_{vJ}^\pm$, the variation of the dimensionless ratio $\omega_1/\omega_2$ is given by
\begin{equation*}
\frac{\delta\left(\omega_1/\omega_2\right)}{\left(\omega_1/\omega_2\right)}
=\frac{\delta\omega_1}{\omega_1}-\frac{\delta\omega_2}{\omega_2}
\cong\left(\frac{1}{\omega_1}-\frac{1}{\omega_2}\right)2A_e\left(\frac{\delta\alpha}{\alpha}+\frac{1}{4}\frac{\delta\mu}{\mu}\right).
\end{equation*}
The sensitivity to variations in $\alpha\mu^{1/4}$ is maximized by selecting transitions which have small values of $\omega_1$ and $\omega_2$ and which additionally differ in sign.

Alternatively, one may measure the ratios $(|\omega_1|+|\omega_2|)/\omega_r$ and $(|\omega_1|-|\omega_2|)/\omega_r$, where $\omega_r$ is some reference energy. This is similar to what has been done with atomic Dysprosium (in which case $\omega_1$ and $\omega_2$ are defined by different isotopic species) \cite{CinLapNgu07,FerCinLap07}. Noting that in the present case $\omega_1+\omega_2$ is sensitive to variations in $\alpha\mu^{1/4}$ whereas $\omega_1-\omega_2$ is not, it follows that whether or not $|\omega_1|\pm|\omega_2|$ is sensitive to variations in $\alpha\mu^{1/4}$ depends on the relative signs of $\omega_1$ and $\omega_2$. In particular, given opposite signs of $\omega_1$ and $\omega_2$ the difference $|\omega_1|-|\omega_2|$ is sensitive to variations in $\alpha\mu^{1/4}$; an immediate benefit of this choice is that systematic effects (e.g., constant frequency shifts) can presumably be controlled or eliminated to a large extent by taking the difference.
Dependence of the reference energy on the variation of the fundamental constants may be neglected if there is no relative enhancement (cancellation of different contributions) there. This is the case for the Cs hyperfine standard and any other hyperfine transition (calculated in Ref.~\cite{FlaTed06}) and practically for any other transition in SiBr.

The preceding expressions of this section are approximate relations. In the remainder of this section, we present more precise formulae. In particular, we include contributions from variations in the additional spectroscopic constants $\omega_ex_e$, $\alpha_{Ae}$, and $B_e$ and, further, we use \emph{ab initio} calculations to determine the precise dependence of the variations of the constants with variations of $\alpha$. En route to calculating the variations of the constants with respect to $\alpha$, we obtain values for the constants themselves, which are presented in Table~\ref{Tab:specconsts} alongside the experimental data. We note an impressive agreement between our computed constants and the experimental constants. In particular, the primary constants $A_e$ and $\omega_e$ agree to better than $1\%$, and we feel that this is indicative of the accuracy of our computed variations of the constants with respect to $\alpha$ presented below. A brief discussion of the computational method is provided in the following section.

From our calculations we obtain the following relations for variations of the constants with respect to variations of $\alpha$ only
\begin{eqnarray}
\frac{\delta A_e}{A_e}&=&2.019\frac{\delta\alpha}{\alpha},\nonumber\\
\frac{\delta \alpha_{Ae}}{\alpha_{Ae}}&=&1.927\frac{\delta\alpha}{\alpha},\nonumber\\
\frac{\delta \omega_e}{\omega_e}&=&-4.5\times10^{-3}\frac{\delta\alpha}{\alpha},\label{Eq:varalpha}
\end{eqnarray}
and variations in constants $\omega_ex_e$ and $B_e$ with respect to variations in $\alpha$ give negligible contribution.

For variations with respect to $\mu$, we make use of analytical formulae, using the appropriate dependence of the constants on the reduced mass. In particular, we have the following relations for variations of the constants with respect to variations of $\mu$ only
\begin{eqnarray}
\frac{\delta \omega_e}{\omega_e}&=&-\frac{1}{2}\frac{\delta\mu}{\mu},\nonumber\\
\frac{\delta \alpha_{Ae}}{\alpha_{Ae}}&=&-\frac{1}{2}\frac{\delta\mu}{\mu},\nonumber\\
\frac{\delta \left(\omega_ex_e\right)}{\omega_ex_e}&=&-\frac{\delta\mu}{\mu},\nonumber\\
\frac{\delta B_e}{B_e}&=&-\frac{\delta\mu}{\mu},\label{Eq:varmu}
\end{eqnarray}
and no variation in $A_e$ with respect to variations in $\mu$.

We consider as an example the two transition energies $\omega_1=\Delta E_{1,5/2}^+$ and $\omega_2=\Delta E_{1,3/2}^+$ (i.e., the transitions identified in Fig.~\ref{Fig:rovibronic}(b)). Using Eqs.~(\ref{Eq:varalpha},\ref{Eq:varmu}), we find the variations in these transition energies to be
\begin{equation*}
\delta \omega_1=\delta \omega_2=\left(851~\mathrm{cm}^{-1}\right)\left(\frac{\delta\alpha}{\alpha}+0.247\frac{\delta\mu}{\mu}\right).
\end{equation*}
The deviation of this expression from the less sophisticated expression, Eq.~(\ref{Eq:dwest}), is small and on the order of the accuracy of the computations. Note, however, that for transitions associated with a higher vibrational quantum number $v$ the deviations from Eq.~(\ref{Eq:dwest}) become more pronounced. For example, suppose the transition $\omega=\Delta E_{24,55.5}^-$ is found to be a convenient transition to probe (with Eqs.~(\ref{Eq:DGv},\ref{Eq:DFJ},\ref{Eq:DEvJ}), $\omega=-0.01~\mathrm{cm}^{-1}$); for this transition, we obtain
\begin{equation*}
\delta\omega=\left(764~\mathrm{cm}^{-1}\right)\left(\frac{\delta\alpha}{\alpha}+0.220\frac{\delta\mu}{\mu}\right).
\end{equation*}

\section{Computational details of $X~{^2\Pi_r}$ rovibronic spectrum}
Fine structure splitting is an inherently relativistic effect; hence, all
the energy calculations employed the relativistic four-component
molecular Dirac-Coulomb Hamiltonian,

\begin{equation*}
H_{\mathrm{DC}}=\sum_{i}h_{\mathrm{D}}(i)+\sum_{i<j}\frac{1}{r_{ij}},
\end{equation*}
where
\begin{equation*}
h_{\mathrm{D}}(i)=c\mathbf{\alpha }_{i}\cdot \mathbf{p}_{i}+\beta
_{i}c^{2}+V_{\mathrm{nuc}}(i).
\end{equation*}
Here, $h_{\mathrm{D}}$ is the one-electron Dirac Hamiltonian, with $V_{\mathrm{nuc}}$ the
nuclear attraction operator for the two nuclei considered, and takes into account the finite nucleus
effect, $\mathbf{\alpha}$ and $\beta $ the four-dimensional Dirac matrices, and the
term $\sum_{i<j}1/r_{ij}$ represents the repulsive Coulomb interaction between electrons.

As SiBr is an open shell system with a single valence electron outside the closed
shell, we employed Fock space coupled cluster (FSCC) method with sectors
(0,0) and (0,1) to account for electron correlation, where
the closed-shell cation served as reference, and an electron was added in
the (0,1) sector, with the model space composed of both $^{2}\Pi _{1/2}$ and
$^{2}\Pi _{3/2}$ molecular orbitals to obtain the potential\ energy curves
for the two states of interest.

An uncontracted aug-cc-pVTZ basis set was used for both atoms \cite%
{WooDun93,WilWooPet99}; 37 electrons were correlated and virtual orbitals with
energies above 35 a.u.~were omitted. All the energy calculations were
performed using the DIRAC program package \cite{DIRAC08}, and the
spectroscopic constants were obtained from the potential energy curves
by solving the rotational-vibrational Schr\"odinger equation numerically using
the program VIBROT \cite{KarLinMal03}. The results are presented and compared
with the experimental values in Table I.

In order to assess the dependence of the values of interest, $A_{e}$, $\omega _{e}$, $\alpha _{A_{e}}$, $\omega _{e}x_e$ and $B_e$, on the fine structure constant, the
calculations were carried out for different values of $x\equiv (\alpha/\alpha_0)^2-1$, and the
derivatives were obtained using numerical differentiation.

\section{Computed Linewidths of Rovibronic Levels}

The natural linewidths of the rovibronic levels place a fundamental limit on the accuracy to which a given transition frequency can be measured, and, therefore, also on the accuracy to which any variation in $\alpha\mu^{1/4}$ may be measured. In Ref.~\cite{FlaKoz07}, rough estimates of the linewidths of the rovibronic levels were obtained for $\Pi$-doublets; here we provide computed values for the SiBr molecule.

To obtain the linewidths we require dipole matrix elements. We begin with the wave function $\Psi_{\gamma v}(\overline{q},R)$ representing the Born-Oppenheimer molecular solutions for a non-rotating molecule (see, e.g., Ref.~\cite{Kro75}),
\begin{equation*}
\Psi_{\gamma v}(\overline{q},R)=\psi_\gamma(\overline{q},R)\phi_{\gamma v}(R),
\end{equation*}
where $R$ is the nuclear separation and $\overline{q}$ encapsulates all electronic space and spin coordinates. Here $\psi_\gamma(\overline{q},R)$ represents the electronic eigensolutions for a given nuclear separation $R$, and $\phi_{\gamma v}(R)$ represents the subsequent eigensolutions for the nuclear vibrational motion. The solutions are assumed to be orthogonal and normalized over the appropriate space, i.e.,
\begin{eqnarray*}
\int\psi_{\gamma^\prime}^*(\overline{q},R)\psi_\gamma(\overline{q},R)d\overline{q}&=&\delta_{\gamma^\prime\gamma},\\
\int\phi_{\gamma v^\prime}^*(R)\phi_{\gamma v}(R)dR&=&\delta_{v^\prime v}.
\end{eqnarray*}

We may write the dipole operator in terms of electronic and nuclear contributions,
\begin{equation*}
\mathbf{D}(\overline{q},R)=\mathbf{D}_e(\overline{q})+\mathbf{D}_n(R).
\end{equation*}
Specifically, the electronic contribution $\mathbf{D}_e(\overline{q})$ is given by
\begin{equation*}
\mathbf{D}_e(\overline{q})=-\sum_i\mathbf{r}_i,
\end{equation*}
where $\mathbf{r}_i$ is the position vector of the $i$-th electron ($\mathbf{r}_i\in\overline{q}$) and the summation runs over all electrons. The nuclear contribution $\mathbf{D}_n(R)$ is
\begin{equation*}
\mathbf{D}_n(R)=\left(\frac{Z_B}{M_B}-\frac{Z_A}{M_A}\right)M_\mathrm{red}R\hat{\mathbf{e}}_z,
\end{equation*}
where $M_i$ and $Z_i$ are the mass and atomic numbers of the nuclei $i=A,B$ and we have assumed the coordinate origin to be at the center of mass with $z$-axis aligned with the internuclear axis. A dipole matrix element between the Born-Oppenheimer wave functions reads
\begin{eqnarray}
\langle \gamma^\prime v^\prime|\mathbf{D}|\gamma v\rangle&=&\iint\psi_{\gamma^\prime}^*(\overline{q},R)\phi_{\gamma^\prime v^\prime}^*(R)
\left[\mathbf{D}_e(\overline{q})+\mathbf{D}_n(R)\right]\nonumber\\
&&\times\psi_\gamma(\overline{q},R)\phi_{\gamma v}(R)d\overline{q}dR\nonumber\\
&=&\int\phi_{\gamma^\prime v^\prime}^*(R)\langle\gamma^\prime|\mathbf{D}|\gamma\rangle\phi_{\gamma v}(R)dR,
\label{Eq:dipolenorot}
\end{eqnarray}
where the $R$-dependent matrix element $\langle\gamma^\prime|\mathbf{D}|\gamma\rangle$ represents the dipole matrix element for a given ``clamped'' nuclear separation $R$ and is given by
\begin{equation*}
\langle\gamma^\prime|\mathbf{D}|\gamma\rangle\equiv\delta_{\gamma^\prime\gamma}\mathbf{D}_n(R)
+\int\psi_{\gamma^\prime}^*(\overline{q},R)\mathbf{D}_e(\overline{q})\psi_\gamma(\overline{q},R)d\overline{q}.
\end{equation*}

We are interested in dipole matrix elements between the vibronic states of a (Hund's case $a$) electronic multiplet. The relevant vibrational wave functions $\phi_{\gamma v}(R)$ are largely independent of the particular multiplet level (i.e., $\phi_{\gamma v}(R)\equiv\phi_{v}(R)$). 
Furthermore, the wave functions $\phi_{v}(R)$ are significant in magnitude only within a small region about the equilibrium nuclear separation $R=R_e$ (note that the Born-Oppenheimer approximation is rooted in the assumption that $\phi_{v}(R)$ varies more rapidly with $R$ than $\psi_\gamma(\overline{q},R)$). As such, we may expand $\langle\gamma^\prime|\mathbf{D}|\gamma\rangle$ in the integrand of Eq.~(\ref{Eq:dipolenorot}) about $R_e$; explicitly to the first order in $(R-R_e)$ this is
\begin{equation*}
\langle\gamma^\prime|\mathbf{D}|\gamma\rangle\cong\langle\gamma^\prime|\mathbf{D}|\gamma\rangle\Big|_{R_e}
+\left.\frac{d\langle\gamma^\prime|\mathbf{D}|\gamma\rangle}{dR}\right|_{R_e}(R-R_e).
\end{equation*}
Subsequent evaluation of the integral gives
\begin{eqnarray}
\langle \gamma^\prime v^\prime|\mathbf{D}|\gamma v\rangle&\cong&\delta_{v^\prime v}\langle\gamma^\prime|\mathbf{D}|\gamma\rangle\Big|_{R_e}
+\left.\frac{d\langle\gamma^\prime|\mathbf{D}|\gamma\rangle}{dR}\right|_{R_e}
\nonumber\\&&
\times R_e\sqrt{\frac{B_e}{\omega_e}}\left(\sqrt{v}\delta_{v^\prime+1,v}+\sqrt{v^\prime}\delta_{v^\prime-1,v}\right),
\nonumber\\&&
\label{Eq:dipole}
\end{eqnarray}
where we have assumed the vibrational wave functions $\phi_{v}(R)$ to be harmonic oscillator eigenfunctions.

The contribution to the natural linewidth for a given decay channel $\gamma v\rightarrow\gamma^\prime v^\prime$ is given by
\begin{equation}
\Gamma(\gamma v\rightarrow\gamma^\prime v^\prime)=\frac{4\omega_{\gamma v,\gamma^\prime v^\prime}^3}{3c^3}\left|\langle \gamma^\prime v^\prime|\mathbf{D}|\gamma v\rangle\right|^2,
\label{Eq:natlinewid}
\end{equation}
where $\omega_{\gamma v,\gamma^\prime v^\prime}$ is the energy difference between the initial and final state, and we have summed over final rotational states. We neglect decay channels within a given vibronic level and between quasi-degenerate vibronic levels for which the energy difference is small.

We begin by considering vibrational decay. Using the MOLPRO computational package \cite{MOLPRO}, we calculated the diagonal matrix elements $\langle {^2\Pi_{1/2}}|\mathbf{D}|{^2\Pi_{1/2}}\rangle$ and $\langle {^2\Pi_{3/2}}|\mathbf{D}|{^2\Pi_{3/2}}\rangle$ at multiple nuclear separation distances $R$ within the vicinity of $R_e$. With numerical differentiation and taking the experimental ratio $\sqrt{B_e/\omega_e}=0.020$, we obtain the results
\begin{equation}
\left.\begin{array}{r}
\left|\langle {^2\Pi_{1/2}},v-1|\mathbf{D}|{^2\Pi_{1/2}},v\rangle\right|\\
\\
\left|\langle {^2\Pi_{3/2}},v-1|\mathbf{D}|{^2\Pi_{3/2}},v\rangle\right|
\end{array}
\right\}
=0.12\sqrt{v}~\mathrm{a.u.}
\label{Eq:vibdipole}
\end{equation}

The decay channel ${^2\Pi_{3/2}},v\rightarrow{^2\Pi_{1/2}},v$ is forbidden in the non-relativistic limit, though it is opened up by spin-orbit mixing of the ${^2\Pi_{1/2}}$ state with the excited electronic ${^2\Sigma_{1/2}}$ state. Due to its purely relativistic origin, the corresponding dipole matrix element proves very difficult to obtain by computational methods to any degree of reliability. To circumvent the need for a direct computational value, we write the dipole matrix element as in Ref.~\cite{FlaKoz07},
\begin{equation*}
\left|\langle {^2\Pi_{3/2}}|\mathbf{D}|{^2\Pi_{1/2}}\rangle\right|\cong\xi\left|\langle {^2\Pi_{3/2}}|\mathbf{D}|{^2\Sigma_{1/2}}\rangle\right|,
\end{equation*}
where $\xi$ is the small parameter quantifying the spin-orbit mixing. We find from computation
$\left|\langle {^2\Pi_{3/2}}|\mathbf{D}|{^2\Sigma_{1/2}}\rangle\right|\!\sim\!0.1~\mathrm{a.u.}$, indicating that $\left|\langle {^2\Pi_{3/2}}|\mathbf{D}|{^2\Pi_{1/2}}\rangle\right|$ is appreciably smaller than the dipole matrix elements in Eqs.~(\ref{Eq:vibdipole}). Thus we conclude that neglect of the decay channel ${^2\Pi_{3/2}},v\rightarrow{^2\Pi_{1/2}},v$ is acceptable as its contribution to linewidths will be overshadowed by the contribution from the vibrational decay.

As a specific example, we consider the natural linewidth of the ${^2\Pi_{1/2}},v=1$ vibronic level. With Eqs.~(\ref{Eq:natlinewid}), we find
\begin{equation*}
\Gamma({^2\Pi_{1/2}},v=1)=5.3\times10^{-17}~\mathrm{a.u.}=0.35~\mathrm{Hz}.
\end{equation*}
We note that this is an order of magnitude larger than the estimate given in Ref.~\cite{FlaKoz07}.

%
%

\section{Conclusion}
Here we have extended upon the Flambaum and Kozlov's work \cite{FlaKoz07} by considering properties of silicon monobromide that make it a prospective candidate for detecting variations in the fine-structure constant $\alpha$ and the proton-to-electron mass ratio $\mu$ (in particular, variations in the combined constant $\alpha\mu^{1/4}$). We have examined the rovibronic spectrum by employing the most accurate experimental data available in the literature, namely that of Bosser~\emph{et al.}~\cite{BosLebRos81}. Furthermore, we present results of \emph{ab initio} calculations for the precise dependence of the spectroscopic constants on variations in $\alpha$. We additionally present calculated values for the natural linewidths of the rovibronic levels which place a fundamental limit on the accuracy to which variations in $\alpha\mu^{1/4}$ may be determined.

As in Ref.~\cite{FlaKoz07}, we emphasize that dedicated measurements are necessary to find precise values for the transition frequencies and determine the best transitions. It is our hope that this work entices experimental progress in this direction.

This work was supported by the Marsden Fund, administered by the Royal Society of New Zealand, and the Australian Research Council. The authors wish to thank Detlev Figgen and Elke Pahl for assistance with computational program packages.

\end{document}